\documentclass[a4paper]{article}
\usepackage{INTERSPEECH2018}

\usepackage{mathtools,calc, epsfig}
\usepackage{amsfonts}
\usepackage{amsopn}
\usepackage{graphicx,caption,lipsum}
\usepackage{booktabs,multirow}
\usepackage{placeins}
\usepackage{xcolor}
\usepackage{enumitem}
\usepackage{xspace,epstopdf,cite,cleveref}
\usepackage{graphicx}
\usepackage{amssymb,amsmath,bm}
\usepackage{textcomp}
\usepackage{mathrsfs}  
\usepackage{xspace}  

\def\vec#1{\ensuremath{\bm{{#1}}}}

\newcommand{\etal}{\textit{et al.\@}}
\newcommand{\etc}{\textit{etc.\@}}
\newcommand{\ie}{\textit{i.e.,\ }}
\newcommand{\vs}{\textit{vs.\ }}

\newcommand{\nedF}{\textit{Neural Entrainment Distance}\xspace} 
\newcommand{\ned}{NED\xspace}

\graphicspath{ {} }

\title{Towards an Unsupervised Entrainment Distance in Conversational Speech using Deep Neural Networks}

\name{\em Md Nasir$^1$, Brian Baucom$^2$, Shrikanth Narayanan$^1$, Panayiotis Georgiou$^1$ }

\address{
  $^1$University of Southern California, Los Angeles, CA, USA\\
  $^2$University of Utah, Salt Lake City, UT, USA}
\email{mdnasir@usc.edu, brian.baucom@utah.edu, \{shri, georgiou\}@sipi.usc.edu}

\begin{document}

\maketitle
\begin{abstract}


Entrainment is a known adaptation mechanism that causes interaction participants to adapt or synchronize their acoustic characteristics.
Understanding how interlocutors tend to adapt to each other's speaking style through entrainment involves measuring a range of acoustic features and comparing those via multiple signal comparison methods.
In this work, we present a turn-level distance measure obtained in an unsupervised manner using a Deep Neural Network~(DNN) model, which we call \nedF (\ned).
This metric establishes a framework that learns an embedding from the population-wide entrainment in an unlabeled training corpus.
We use the framework for a set of acoustic features and validate the measure experimentally by showing its efficacy in distinguishing real conversations from fake ones created by randomly shuffling speaker turns.
Moreover, we show real world evidence of the validity of the proposed measure.
We find that high value of \ned is associated with high ratings of emotional bond in suicide assessment interviews, which is consistent with prior studies.

\end{abstract}

\noindent\textbf{Index Terms}: entrainment, deep neural network, unsupervised learning, embeddings, behavioral analysis, conversational speech

\section{Introduction}
Vocal entrainment is an established social adaptation mechanism.
It can be loosely defined as one speaker's spontaneous adaptation to the speaking style of the other speaker.
Entrainment is a fairly complex multifaceted process and closely associated with many other mechanisms such as  coordination, synchrony, convergence \etc~ 
While there are various aspects and levels of entrainment~\cite{levitan2011measuring}, there is also a general agreement that entrainment is a sign of positive behavior towards the other speaker~\cite{welkowitz1973interrelationships, hirschberg2011speaking, bevnuvs2014social}.
High degree of vocal entrainment has been associated with various interpersonal behavioral attributes, such as high empathy~\cite{xiao2013modeling}, more agreement and less blame towards the partner and positive outcomes in couple therapy~\cite{nasir2016complexity}, and high emotional bond~\cite{nasir2017complexity}.
A good understanding of entrainment provides insights to various interpersonal behaviors and facilitates the recognition and estimation of these behaviors in the realm of Behavioral Signal Processing~\cite{georgiou2011_behavioral-sign,narayanan2013behavioral}.
Moreover, it also contributes to the modeling and development of `human-like' spoken dialog systems or conversational agents.

Unfortunately, quantifying entrainment has always been a challenging problem. 
There is a scarcity of reliable labeled speech databases on entrainment, possibly due to the subjective and diverse nature of its definition.
This makes it difficult to capture entrainment using supervised models, unlike many other behaviors.
Early studies on entrainment relied on highly subjective and context-dependent manual observation coding for measuring entrainment.
The objective methods based on extracted speech features employed classical synchrony measures such as Pearson's correlation~\cite{levitan2011measuring} and traditional (linear) time series analysis techniques\cite{kousidis2009convergence}.
Lee \etal~\cite{lee2014computing, xiao2013modeling} proposed a measure based on PCA representation of prosody and MFCC features of consecutive turns.
Most of the these approaches assume a linear relationship between features of consecutive speaker turns which is not necessarily true, given the complex nature of entrainment. 
For example, the effect of rising pitch or energy can potentially have a nonlinear influence across speakers.

Recently, various complexity measures (such as largest Lyapunov exponent) of feature streams based on  nonlinear dynamical systems modeling showed promising results in capturing entrainment~\cite{nasir2016complexity,nasir2017complexity}.
A limitation of this modeling, however, is the assumption of the short-term stationary or slowly varying nature of the features.
While this can be reasonable for global or session-level complexity, the measure is not very meaningful capturing turn-level or local entrainment.
Nonlinear dynamical measures also suffer from scalability to a multidimensional feature set, including spectral coefficients such as MFCCs. 
Further, all of the above metrics are knowledge-driven and do not exploit the vast amount of information that can be gained from existing interactions.

A more holistic approach is to capture entrainment in consecutive speaker turns through a more robust nonlinear function.
Conceptually speaking, such a formulation of entrainment is closely related to the problem of learning a transfer function which maps vocal patterns of one speaker turn to the next.
A compelling choice to nonlinearly approximate the transfer function would be to employ Deep Neural Networks~(DNNs).
This is supported by  recent promising applications of deep learning models, both in supervised and unsupervised paradigm, in modeling and classification of emotions and behaviors from speech.
For example in \cite{li2017_unsupervised-la} the authors learned, in an unsupervised manner, a latent embedding towards identifying behavior in out-of-domain tasks.
Similarly in \cite{jati2018_neural-predicti,jati2017_speaker2vec:-un} the authors employ Neural Predictive Coding to derive embeddings that link to speaker characteristics in an unsupervised manner.

We propose an \emph{unsupervised training} framework to \emph{contextually} learn the transfer function that ties the two speakers.
The learned bottleneck embedding contains cross-speaker information closely related to entrainment.
We define a distance measure between the consecutive speaker turns represented in the bottleneck feature embedding space.
We call this metric the \textbf{\nedF (\ned)}.

Towards this modeling approach we use features that have already been established as useful for entrainment.
The majority of research \cite{levitan2011measuring,lee2010quantification, lee2014computing, nasir2016complexity, nasir2017complexity} focused on  prosodic features like pitch, energy, and speech rate.
Others also analyzed entrainment in spectral and voice quality features~\cite{lee2014computing, xiao2013modeling}.
Unlike classical nonlinear measures, we jointly learn from a \emph{multidimensional feature set} comprising of prosodic, spectral, and voice quality features.

We then experimentally investigate the validity and effectiveness of the \ned measure in association with interpersonal behavior.

\section{Datasets}
We use two datasets in this work: the training is done on the Fisher Corpus English Part 1 (LDC2004S13)~\cite{cieri2004fisher} and testing on the Suicide Risk Assessment corpus~\cite{bryan2017emotional}, along with Fisher.
\begin{itemize}
\item \textbf{Fisher Corpus English Part 1}: It consists of spontaneous telephonic dyadic conversations between native English speakers.
There are 5850 such conversations, each lasting up to 10 minutes.
The manual transcripts of the corpus contain time-stamps of speaker turn boundaries as well as boundaries of pauses within a turn.

\item \textbf{Suicide Risk Assessment corpus}: This dataset contains recorded conversations of active duty military personnel with their therapist during suicide risk assessment sessions.
The participants were suicidal patients who either had attempted suicide or had suicidal thoughts prior to the sessions.
The subset of the corpus employed in the current work included therapist-patient interviews of 54 subjects, each session ranging from 10  minutes to 1 hour. 
They were asked questions related to their personal history, reasons leading to their suicidal ideations, elaboration of their reasons for living \etc 
~~Immediately after the interview sessions, the patient was asked to provide with a self-reported score for perceived \textit{emotional bond}, an attribute which entails the therapist's empathy for the patient and the patient's feeling of trust towards them. It was rated on a scale from 1 to 10.

\end{itemize}

\section{Modeling of Neural Entrainment Distance}

\subsection{Preprocessing}
A number of audio preprocessing steps are required in the entrainment framework for obtaining boundaries of relevant segments of audio from consecutive turns.
First, we perform voice activity detection~(VAD) to identify the speech regions.
Following this, speaker diarization is performed in order to distinguish speech segments spoken by different speakers.
However, our training dataset, the Fisher corpus also contains transcripts with speaker turn boundaries as well as timings for pauses within a turn.
Since, these time stamps appeared to be reasonably accurate, we use them as oracle VAD and diarization.
On the other hand, for the Suicide Risk Assessment corpus, we perform  VAD and diarization on raw audio to obtain the turn boundaries.
Subsequently, we also split a single turn into inter-pausal units (IPUs) if there is any pause of at least 50 ms present within the turn.
For the purpose of capturing entrainment-related information, we only consider the initial and the final IPU of every turn.
This is done based on the hypothesis that during a turn-taking, entrainment is mostly prominent between the most recent IPU of previous speaker's turn and the first IPU of the next speaker's turn~\cite{levitan2011measuring}.

\subsection{Feature Extraction}
We extract 38 different acoustic features from the segments (IPUs) of our interest. 
The extracted feature set includes 4 prosody features (pitch, energy and their first order deltas), 31 spectral features~(15 MFCCs, 8 MFBs, 8 LSFs) and 3 voice quality features~(shimmer and 2 variants of jitter).
We found in our early analysis that derivatives of spectral and voice quality features do not seem to contribute significantly to entrainment\footnote{These features showed very low correlation~($\rho<0.05$) across consecutive turns in our initial analysis} and hence we do include them for the \ned model.
The feature extraction is performed with a Hamming window of 25 ms width and 10 ms shift using the OpenSMILE toolkit~\cite{eyben2010opensmile}.
For pitch, we perform an additional post-processing by applying a median-filter based smoothing technique~(with a window size of 5 frames) as pitch extraction is not very robust and often prone to errors, such as halving or doubling errors.
We also perform z-score normalization of the features across the whole session, except for pitch and energy features, which are normalized by dividing them by their respective means. 
\begin{figure}[h]
  \centering
  \vspace{-0.5cm}  
  \includegraphics[trim= 4cm 2cm 1cm 1cm, clip,width=1.0\columnwidth]{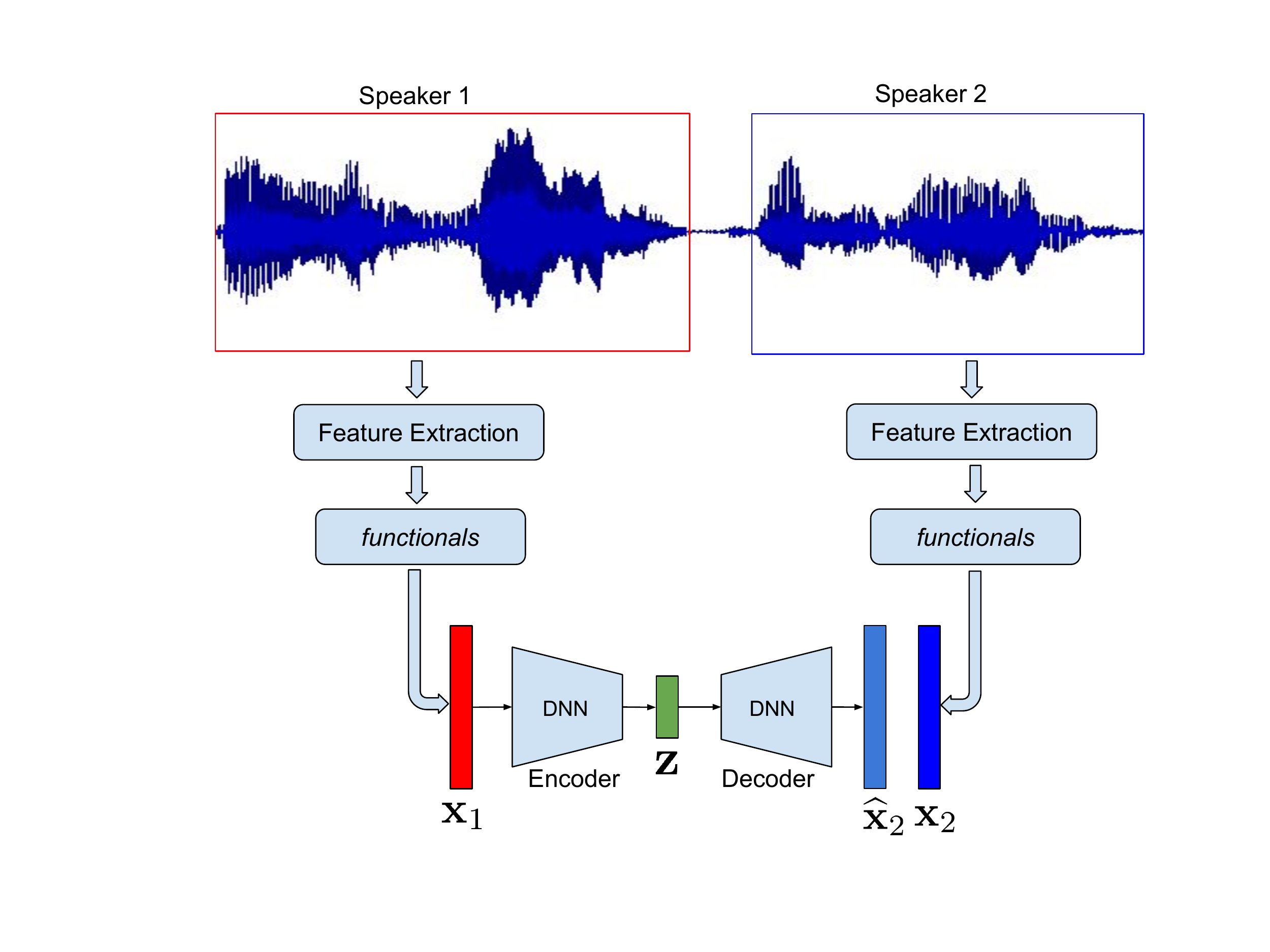}
  \caption{An overview of unsupervised training of the model}
  
\vspace{-0.4cm}
  \label{fig:model}
\end{figure}

\subsection{Turn-level Features}
We propose to calculate NED as directional entrainment-related measure from speaker 1 to speaker 2 for a change of turn as shown in Figure \ref{fig:model}.
The segments of interest in this case are the final IPU of speaker 1's turn  and the initial IPU of the subsequent turn by speaker 2, marked by the bounding boxes in the figure.
As turn-level features, we compute six statistical functionals over all frames in those two IPUs, generating two sets of functionals of features for each pair of turns.
The functionals we compute are as follows: mean, median, standard deviation, 1st percentile, 99th percentile and range between 99th and 1st percentile.
Thus we obtain $38 \times 6=228$ turn-level features from each IPU representing the turn.
Let us denote the turn-level feature vector of the final IPU of speaker 1 and the initial IPU of speaker 2 as $\mathbf{x}_1$ and $\mathbf{x}_2$, respectively, for further discussion in the paper.

\subsection{Modeling with Neural Network}
Most work in the entrainment literature directly computes a measure between $\mathbf{x}_1$ and $\mathbf{x}_2$ (such as correlation~\cite{levitan2011measuring}) or their lower-dimensional representations~\cite{lee2014computing}.
However, one conceptual limitation of all these approaches is that turn-level features $\mathbf{x}_1$ and $\mathbf{x}_2$ do not only contain the underlying acoustic information that can be entrained across turns, but also speaker-specific, phonetic and paralinguistic information that is specific to the corresponding turns and not influenced by the previous turn~(non-entrainable).
If we represent those two types of information as vector embeddings, $\mathbf{e}$ and $\mathbf{q}$ respectively, we can model turn-level feature vectors $\mathbf{x}$ as a nonlinear function $\mathscr{F}(\cdot)$ over them, \ie$\mathbf{x}_1 = \mathscr{F}(\mathbf{e}_1,\mathbf{q}_1)$ and $\mathbf{x}_2 = \mathscr{F}(\mathbf{e}_2,\mathbf{q}_2)$.
In this formulation, the distance between $\mathbf{e}_1$ and $\mathbf{e}_2$ should be zero in the  hypothetical case of `perfect' entrainment. 

Our goal is to approximate the inverse mappings that maps the feature vector  $\mathbf{x}$ to entrainment embedding $\mathbf{e}$ and ideally to learn the same from `perfect' or very highly entrained turns.
Unfortunately, in absence of such a dataset, we learn it from consecutive turns in real data where entrainment is present, at least to some extent.
As shown in Figure \ref{fig:model}, we adopt a feed-forward deep neural network~(DNN) as an encoder for this purpose.\\

The different components of the model are described below:
\begin{enumerate}
\item First we use $\mathbf{x}_1$ as the input to the encoder network.
We choose the output of the encoder network, $\mathbf{z}$ to be undercomplete representation of $\mathbf{x}_1$, by restricting the dimensionality of $\mathbf{z}$ to be lower than that of $\mathbf{x}$.
\item  $\mathbf{z}$ is then passed through another feed-forward~($\mathbf{z}$) network used as decoder to predict $\mathbf{x}_2$. The output of the decoder is denoted as $\widehat{\mathbf{x}}_2$. 
\item Then $\widehat{\mathbf{x}}_2$ and its reference $\mathbf{x}_2$ are compared to obtain the loss function of the model, $\mathcal{L}(\mathbf{x}_2,\widehat{\mathbf{x}}_2)$.
\end{enumerate}

Even though this deep neural network resembles autoencoder architectures, it does not reconstruct itself but rather tries to encode relevant information from one turn to predict the next turn, parallel to  \cite{jati2018_neural-predicti,jati2017_speaker2vec:-un, li2017_unsupervised-la}.
Thus the bottleneck embedding $\mathbf{z}$  can be considered closely related to the entrainment embedding $\mathbf{e}$ mentioned above.

\subsection{Unsupervised Training of the Model}
In this work, we use two fully connected layers as hidden layers both in the encoder and decoder network.
Batch normalization layers and Rectified Linear Unit~(ReLU) activation layers~(in respective order) are used between fully connected layers in both of the networks.
The dimension of the embedding is chosen to be 30.
The number of neuron units in the hidden layers are: [~228~$\rightarrow$~128~$\rightarrow$~30~$\rightarrow$~128~$\rightarrow$~228~].
We use smooth L1 norm, a variant of L1 norm which is more robust to outliers~\cite{Girshick2015FR}, so that
\vspace{-0.2cm}
\begin{equation}
\mathcal{L}(\mathbf{x}_2,\widehat{\mathbf{x}}_2) = \|\mathbf{x}_2-\widehat{\mathbf{x}}_2\|^{\text{smooth}}_{1} = \sum_{k=1}^{N} \text{smooth}_{L_{1}}(x_{2k} - \widehat{x}_{2k}),
\end{equation} where 
\vspace{-0.6cm}
\begin{equation}
\text{smooth}_{L_{1}}(d) =    \begin{cases}
      0.5 d^2, & \text{if}\ | d  | \leq 1 \\
      | d | - 0.5, & \text{otherwise}
    \end{cases}
\end{equation}
and $N$ is the dimension of $\mathbf{x}$ which is 228 in our case.\\
For training the network, we choose a subset~(80\% of all sessions) of Fisher corpus and use all turn-level feature pairs ($\mathbf{x}_1, \mathbf{x}_2$).
We employ the Adam optimizer~\cite{kingma2014adam} and a minibatch size of 128 for training the network.
The validation error is computed on the validation subset~(10\% of the data) of the Fisher corpus and the best model is chosen.

\subsection{Neural Entrainment Distance~(NED) Measure}
After the unsupervised training phase, we use the encoder network to obtain the embedding representation~($\mathbf{z}$) from any turn-level feature vector $\mathbf{x}$.
To quantify the entrainment from a turn to the subsequent turn, we extract turn-level feature vectors from their final and initial IPUs, respectively, denoted as $\mathbf{x}_i$ and $\mathbf{x}_j$.
Next we encode $\mathbf{x}_i$ and $\mathbf{x}_j$ using the pretrained encoder network and obtain $\mathbf{z}_i$ and $\mathbf{z}_j$ as the outputs, respectively.
Then we compute a distance measure $d_\mathrm{{NE}}$, which we term \nedF (\ned), between the two turns by taking smooth L1 distance $\mathbf{z}_i$ and $\mathbf{z}_j$. 
\vspace{-0.3cm}
\begin{equation}
d_\mathrm{{NE}}(\mathbf{x}_i,\mathbf{x}_j) = \|\mathbf{z}_i-\mathbf{z}_j\|^{\text{smooth}}_{1} = \sum_{k=1}^{M} \text{smooth}_{L_{1}}(z_{ik} - z_{jk}),
\vspace{-0.3cm}
\end{equation}

where $\text{smooth}_{L_{1}}(\cdot)$ is defined in Equation (2) and $M$ is the dimensionality of the embedding.
Note that even though smooth L1 distance is symmetric in nature, our distance measure is still asymmetric because of the directionality in the training of the neural network model.

\section{Experimental Results}
We conduct a number of experiments to validate  \ned as a valid proxy metric for entrainment.

\subsection{Experiment 1: Classification of real \vs fake sessions}
We first create a fake session~($\mathcal{S}_{fake}$) from each real session~($\mathcal{S}_{real}$) by randomly shuffling the speaker turns.
Then we run a simple classification experiment of using the \ned measure to identify the real session from the pair ($\mathcal{S}_{real}$, $\mathcal{S}_{fake}$). The steps of the experiments are as follows:
\begin{enumerate}
\item We compute \ned for each (overlapping) pair of consecutive turns and their average across the session for both sessions in the pair ($\mathcal{S}_{real}$, $\mathcal{S}_{fake}$).
\item The session with lower \ned is inferred to be the real one. 
The hypothesis behind this rule is that higher entrainment is seen across consecutive turns than randomly paired turns and is well captured through a lower value of proposed measure.
\item If the inferred real session is indeed the real one, we consider it to be correctly classified.
\end{enumerate}
We compute classification accuracy averaged over 30 runs (to account for the randomness in creating the fake session) and report it in Table \ref{tab:expt1}.
The experiment is conducted on two datasets: a subset~(10\%) of Fisher corpus set aside as test data and Suicide corpus.
We use a number of baseline measures:
\begin{itemize}
\item \underline{Baseline 1}: smooth L1 distance directly computed between turn-level features~($\mathbf{x}_i$ and $\mathbf{x}_j$)
\item \underline{Baseline 2}: PCA-based symmetric acoustic similarity measure by Lee \etal~\cite{lee2014computing}
\item \underline{Baseline 3}: Nonlinear dynamical systems-based complexity measure~\cite{nasir2017complexity}. 
\vspace{-1ex}
\end{itemize}
For the baselines, we conduct the classification experiments in a similar manner.
Since Baseline 1 and 2 have multiple measures, we choose the best performing one for reporting, thus providing an upper-bound performance.
Also, for  baseline 2 we choose the session with higher value of the measure as real, since it measures similarity.

\begin{table}[h]
   \centering
      \fontsize{10}{10}\selectfont
   \renewcommand{\arraystretch}{1.4}
      \resizebox{0.8\linewidth}{!}{
       \begin{tabular}{c|c|c}
       \hline
       \multirow{2}{*}{Measure}  & \multicolumn{2}{c}{Classification accuracy~(\%)} \\
		\cline{2-3}
        & Fisher corpus  & Suicide corpus \\

       \hline
		Baseline 1  &  $72.10~(5.83)$     &  $70.44~(6.69)$             \\      
        Baseline 2  &  $92.32~(3.01)$     &   $88.12~(5.93)$          \\
        Baseline 3 &    $90.21~(5.40)$  &  $88.54~(5.87)$           \\           
                    \hline  
\ned &  $98.87~(0.97)$     &  $91.92~(2.32)$         \\  
       \hline
    \end{tabular}}
\vspace{0.2cm}
    \caption{Results of Experiment 1: classification accuracy~(\%) of real \textit{vs.} fake sessions (averaged over 30 runs; standard deviation  shown in parentheses)
    \vspace{-1.2cm} \\
}
    \label{tab:expt1}
    \end{table}
 \vspace{-0.1cm}

As we can see in Table \ref{tab:expt1}, our proposed \ned measure achieves higher accuracy than all baselines on the Fisher corpus.
The accuracy of our measure declines in the Suicide corpus as compared to the Fisher corpus, which is probably due to data mismatch as the model was trained on Fisher (mismatch of  acoustics, recording conditions, sampling frequency, interaction style \etc).
However, our measure still performs better than all baselines on Suicide corpus.

\vspace{-0.3cm}
\subsection{Experiment 2: Correlation with Emotional Bond}
According to prior work, both from domain theory~\cite{bryan2017emotional} and from experimental validation~\cite{nasir2017complexity}, a high emotional bond in patient-therapist interactions in the suicide therapy domain is associated with more entrainment.
In this experiment, we compute the correlation of the proposed \ned measure with the patient-perceived emotional bond ratings.
Since the proposed measure is asymmetric in nature, we compute the measures for both patient-to-therapist and therapist-to-patient entrainment. 
We also compute the correlation of emotional bond with the baselines used in Experiment 1.
We report Pearson's correlation coefficients ($\rho$) for this experiment in Table \ref{tab:emo} along with their $p$-values. 
We test against the null hypothesis $H_0$ that there is no linear association between emotional bond and the candidate measure.

\begin{table}[b]
   \centering
      \fontsize{12}{12}\selectfont
   \renewcommand{\arraystretch}{1.4}
      \resizebox{0.6\linewidth}{!}{
       \begin{tabular}{c|c|c}
       \hline
       \multirow{2}{*}{Measure}  & \multicolumn{2}{c}{Pearson's correlation} \\
		\cline{2-3}
        & $\rho$  & $p$-$\mathrm{value}^{*}$ \\

       \hline
		Baseline 1  &   $-0.1980$     &  $0.2031$            \\      
        Baseline 2  &    $0.2480$     &  $0.1132$          \\
        Baseline 3 &    $-0.3815$     &  $0.0127$           \\           
                    \hline  
\ned-TP &  $-0.1317$     &  $0.3999$       \\                      
\ned-PT &  $-0.4479$     &  $0.0095$         \\  
       \hline
    \end{tabular}}
\vspace{0.2cm}
    \caption{Correlation between emotional bond and  various measures;   
TP: therapist-to-patient, PT: patient-to-therapist
    \vspace{0.1cm} \\
    \footnotesize	   \emph{${}^* p<0.05$ indicates statistically significant~(strong) correlation}}
    \label{tab:emo}
    \end{table}
Results in Table \ref{tab:emo} show that the patient-to-therapist \ned is negatively correlated with emotional bond with high statistical significance~($p<0.01$).
This negative sign is consistent with previous studies as higher distance in acoustic features indicates lower entrainment. 
However, the therapist-to-patient \ned does not have a significant correlation with emotional bond.
A possible explanation for this finding is that the emotional bond is reported by the patient and influenced by the degree of their perceived therapist-entrainment.
Thus, equipped with an asymmetric measure, we are also able to identify the latent directionality of the emotional bond metric.
The complexity measure (Baseline 2) also shows statistically significant correlation, but the value of $\rho$ is lower than that of the proposed measure.

\begin{figure}[t]
  \centering
  \vspace{-0.3cm}  
  \includegraphics[trim= 2.5cm 8.3cm 0cm 7.7cm, clip,width=1.0\columnwidth]{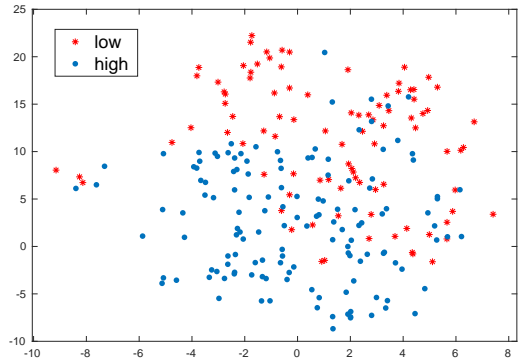}
  \caption{t-SNE plot of difference vector of encoded turn-level embeddings for sessions with low and high emotional bond}
  
\vspace{-0.4cm}
  \label{fig:tsne}
\end{figure}

To analyze the embeddings encoded by our model, we also compute a t-SNE~\cite{maaten2008visualizing} transformation of the difference of all patient-to-therapist turn embedding pairs, denoted as $\mathbf{z}_i-\mathbf{z}_j$ in Equation (3).
Figure \ref{fig:tsne} shows the results of a session with high emotional bond and another one with low emotional bond~(with values of 7 and 1 respectively) as a 2-dimensional scatter plot.
Visibly there is some separation between the sessions with low and high emotional bond.

\section{Conclusion and Future Work}
In this work, a novel deep neural network-based \nedF~(\ned) measure is proposed for capturing entrainment in conversational speech. 
The neural network architecture consisting of an encoder and a decoder is trained on the Fisher corpus in an unsupervised training framework and then the measure is defined on the bottleneck embedding.
We show that the proposed measure can distinguish between real and fake sessions by capturing presence of entrainment in real sessions. 
In this way we also validate the natural occurrence of vocal entrainment in dyadic conversations, well-known in psychology literature\cite{watt1996dynamic,andersen1984exchange, burgoon2007interpersonal}.
We further show that the measure for patient-to-therapist direction achieves statistically significant correlation with their perceived emotional bond.
The proposed measure is asymmetric in nature and can be  useful for analyzing different interpersonal~(especially directional) behaviors in many other applications.
Given the benefits shown by the unsupervised data-driven approach we will employ    Recurrent Neural Networks~(RNNs) to better capture temporal dynamics.
We also intend to explore (weakly) supervised learning of entrainment using the bottleneck embeddings as features, in presence of session-level annotations.

\section{Acknowledgements}

The U.S. Army Medical Research Acquisition Activity, 820 Chandler Street, Fort Detrick MD 21702- 5014 is the awarding and administering acquisition office.
This work was supported by the Office of the Assistant Secretary of Defense for Health Affairs through the Military Suicide Research Consortium under Award No. W81XWH-10-2-0181, and through the Psychological Health and Traumatic Brain Injury Research Program under Award No. W81XWH-15-1-0632.
Opinions, interpretations, conclusions and recommendations are those of the author and are not necessarily endorsed by the Department of Defense.

\bibliographystyle{IEEEtran}

\bibliography{dynamic,entrain,tools}

\end{document}